\documentclass{article}

\usepackage{arxiv}

\usepackage[utf8]{inputenc} 
\usepackage[T1]{fontenc}    
\usepackage{hyperref}       
\usepackage{url}            
\usepackage{booktabs}       
\usepackage{amsfonts}       
\usepackage{nicefrac}       
\usepackage{microtype}      
\usepackage{lipsum}		
\usepackage{graphicx}
\usepackage{natbib}
\usepackage{doi}

\title{Dynamic Time-Alignment of Dimensional Annotations of Emotion using Recurrent Neural Networks}

\date{} 					

\author{ 
    Sina Alisamir \\
	\textit{Atos \& LIG}\\
	\textit{Univ. Grenoble Alpes}\\
	Grenoble, France \\
	sina.alisamir@univ-grenoble-alpes.fr \\
	\And
	Fabien Ringeval \\
	\textit{Grenoble INP, LIG}\\
	\textit{Univ. Grenoble Alpes, Inria, CNRS}\\
	Grenoble, France \\
	fabien.ringeval@univ-grenoble-alpes.fr \\
	\And
	François Portet \\
	\textit{Grenoble INP, LIG}\\
	\textit{Univ. Grenoble Alpes, Inria, CNRS}\\
	Grenoble, France \\
	francois.portet@univ-grenoble-alpes.fr \\
}

\renewcommand{\headeright}{}
\renewcommand{\undertitle}{}
\renewcommand{\shorttitle}{Dynamic Time-Alignment of Dimensional Annotations of Emotion}

\hypersetup{
pdftitle={A template for the arxiv style},
pdfsubject={q-bio.NC, q-bio.QM},
pdfauthor={David S.~Hippocampus, Elias D.~Striatum},
pdfkeywords={First keyword, Second keyword, More},
}

\begin{document}
\maketitle

\begin{abstract}
Most automatic emotion recognition systems exploit time-continuous annotations of emotion to provide fine-grained descriptions of spontaneous expressions as observed in real-life interactions. As emotion is rather subjective, its annotation is usually performed by several annotators who provide a \textit{trace} for a given dimension, i.e. a time-continuous series describing a dimension such as arousal or valence. However, annotations of the same expression are rarely consistent between annotators, either in time or in value, which adds bias and delay in the trace that is used to learn predictive models of emotion. We therefore propose a method that can dynamically compensate inconsistencies across annotations and synchronise the traces with the corresponding acoustic features using Recurrent Neural Networks. Experimental evaluations were carried on several emotion data sets that include Chinese, French, German, and Hungarian participants who interacted remotely in either noise-free conditions or in-the-wild. 
The results show that our method can significantly increase inter-annotator agreement, as well as correlation between traces and audio features, for both arousal and valence. In addition, improvements are obtained in the automatic prediction of these dimensions using simple light-weight models, especially for valence in noise-free conditions, and arousal for recordings captured in-the-wild. 
\end{abstract}

\keywords{Affective computing \and Emotion recognition \and Dynamic time-alignment \and Recurrent neural networks}

\section{Introduction}

Automatic detection of apparent human emotions is of growing interest as it has many real-life applications, touching mostly upon education~\cite{Tsatsou18-ALB}, health~\cite{cummins2015review}, and entertainment~\cite{cosentino2018group}. Affective computing exploits psychological theories of emotion that describe expressions of affect with either a categorical, or a dimensional model: categorical representations view emotion as different classes~\cite{Ekman93-FEA}, such as the basic emotions~\cite{Ekman11-WIM}, whereas dimensional representations describe emotion with different scales~\cite{Russell80-ACM}, such as arousal (ranging from active to passive) and valence (ranging from pleasant to unpleasant). Although some attempts have been made to detail the cognitive processes involved in the emotional experience~\cite{Wehrle01-TCM}, 
the arguably dominant approach in modeling affect relies on dimensional representations of emotion, as they allow a fine differentiation of real-life expressions~\cite{schuller2012avec}. 
Such representation of emotion nonetheless implies some well-known issues, such as biases and delays in annotations, which can vary considerably according to the annotators and the peculiarities of the judged expression~\cite{Soroosh13-AAC}. In this paper, our aim is to address this problem.

Methods have been proposed to create a unified \textit{trace} of a set of time-continuous annotations of emotion~\cite{Ringeval18-A2W}, which is usually referred to as ‘Gold Standard’ (GS). One of the main issues in creating a unified GS that can be reliably used for affect modelling stems from annotator reaction delay, which is defined as the time it takes for an annotator to perceive an acoustic event, evaluate it, and report the value best describing the emotional expression according to the chosen scale~\cite{khorram2019jointly}.

When annotator reaction times are taken into account, a significant performance increase can be observed in the automatic recognition of emotion from speech~\cite{Soroosh13-AAC,khorram2019jointly,nicolle2012robust}. Existing methods, however, make the hypothesis that the annotation delay is constant over a whole interaction sequence, whereas it can vary greatly within the same sequence~\cite{mariooryad2014correcting}. In addition, some approaches predict emotions with delays like annotators, which is problematic for conversational systems subject to -- almost -- real-time constraints or systems that use detected speech segments as input, since annotator delay may vary up to several seconds.

\begin{figure*}[t]
  \centering
  \includegraphics[width=\linewidth]{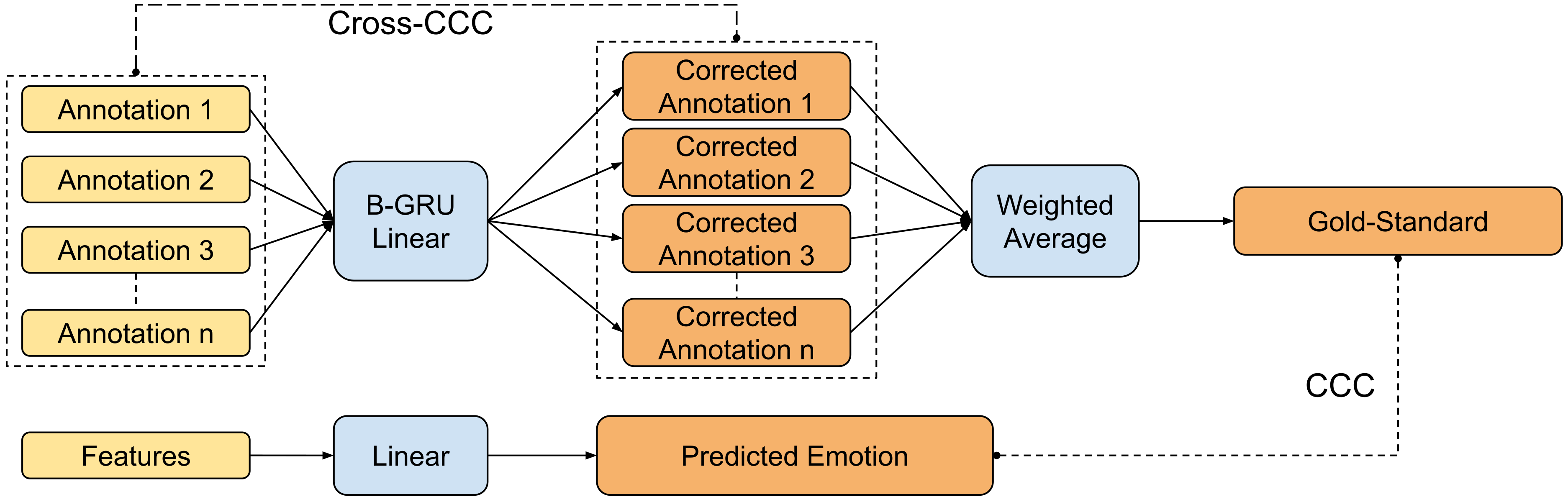}
  \caption{Flowchart of the proposed system for dynamic time-alignment of continuous ratings of apparent emotion with corresponding acoustic events. The system is composed of a first model (B-GRU/Linear layers) that performs corrections to the annotations while preserving their original shape, and of a second model (Linear layer) that predicts the generated GS from the acoustic features. The two models are learned jointly with a loss function based on both CCC and Cross-CCC, where the former accounts for the similarity between the generated GS (weighted average of the corrected annotations) and the emotion predicted from the features, and the latter makes sure that the overall shape of the original annotations is preserved when corrections are applied to them; B-GRU: Bidirectional Gated Recurrent Units.}
  \label{fig:GSgenerator}
\end{figure*}

In this work, we define a tandem of Neural Networks (NNs) that dynamically correct inconsistencies in continuous annotations of emotion, cf. Figure~\ref{fig:GSgenerator}. Whereas a first model has the objective of correcting annotations while preserving their overall shape, another model predicts these corrected annotations from the acoustic features. By jointly learning the two models, the system ultimately provides dynamic corrections of the annotations by aligning them with their corresponding events present in the speech. In effect, the system is forced to provide corrections that match what can be linearly predicted from the data. It is worth to note that our method does not make use of the acoustic features as input of the first model correcting the annotation, as the same features are used to predict emotion.

We performed various evaluations of the generated GS and compared the results with prior methods on two benchmark data sets (RECOLA and SEWA) that include dimensional annotations of emotion for different cultures (Chinese, French, German, Hungarian), and recording conditions (noise-free, in-the-wild). More specifically, we quantified the inter-annotator agreement, the correlation between the GS and the acoustic data, and evaluated the performance obtained in the automatic recognition of emotion from speech, using either the original GS (simple average of the traces), or the one created by our system.

In order to deal with issues related to the fact that features and annotations have different frequency bandwidths~\cite{khorram2017capturing}, we also investigated the interest of a low-pass filter with a cutoff frequency that can be learned for both signals, using a convolutional layer with a Sinc function as kernel. We followed a curriculum in the learning of the model~\cite{bengio2009curriculum}, by starting with a low cutoff frequency, which helps to gradually incorporate more details of the signals during the learning phase.

Results show that: (i) the inter-annotator agreement is either preserved or increased when corrections are applied to the annotations, (ii) corrected annotations have a higher correlation with acoustic features, (iii) the Sinc function helps the system to return predictions with smooth trajectories, and (iv) both the Sinc function and the generated GS can improve the performance obtained in the automatic recognition of emotion from speech. 

\section{Related Work}
\label{sec:related}

As events that trigger changes in the annotation of emotion are present in the data, several studies investigated synchronisation techniques by using audiovisual descriptors to compensate for delays in annotator response. Nicolle et al.~\cite{nicolle2012robust} proposed to maximise the correlation between facial descriptors and annotations to estimate this delay, which is applied as an offset on the features. Other measures were also later exploited, such as mutual information~\cite{mariooryad2014correcting}, or Concordance Correlation Coefficient (CCC)~\cite{li89,he2015multimodal}. A majority vote triplet embedding scheme was then exploited in conjunction with Dynamic Time Warping (DTW) methods to account for inconsistencies in annotations of emotion in ~\cite{booth2018fusing}. Even though DTW methods adjusted the annotations based on a single reference feature, the authors reported improvement in automatic recognition of valence on RECOLA data set.

Long Short-Term Memory (LSTM) Recurrent Neural Networks (RNNs), which can capture long-term temporal dependencies, have been used on individual ratings of emotion to learn dependencies between annotations in a single multi-task problem~\cite{ringeval2015prediction}. 
A downsampling/upsampling network based on Convolutional Neural Networks (CNNs) has also shown the interest of exploiting long-term temporal dependencies for continuous emotion recognition from speech~\cite{khorram2017capturing}. More recently, a Sinc function has been used as the kernel of CNNs to jointly learn annotation delay and emotion~\cite{khorram2019jointly}. While these methods achieved state-of-the-art performance on various benchmark data sets, and showed the value of compensating inconsistencies in annotations of emotion, they did not allow for dynamic annotation correction.

It is worth to mention that a CNN with a Sinc function as kernel (Sinc layer) has been successfully used with two trainable cutoff frequencies in order to learn band-pass filters for automatic speaker and speech recognition~\cite{ravanelli2018speaker, ravanelli2018interpretable}. Authors have shown that more interpretable filters can be obtained with Sinc layers compared to using generic CNNs. Here, our Sinc layer has only one learnable parameter which is the cutoff frequency to smooth the input and output signals. We start the training with a low cutoff frequency, i.e. a highly smoothed signal, so that the model can automatically and gradually adjust the details needed to increase its performance, thus following a curriculum.

\section{Method}
\label{sec:method}

In this section, We describe the method we developed to align continuous annotations of apparent emotion with associated events present in acoustic signal. The system is based on joint learning of a tandem of NNs that are depicted in Figure \ref{fig:GSgenerator}. 

The first model is composed of B-GRU (Bidirectional GRU) followed by a linear layer, and has the objective of correcting annotations while preserving their overall shape. We preferred a bidirectional network use past information when correction annotations. 
Preservation of the original shape of the annotations is obtained by a loss function based on the Cross-CCC, which is computed as the average of the CCCs obtained between delayed versions of the corrected annotations and the original time-series, with steps of 100\,ms and a maximum duration of ten seconds, to cover delays up to this value. 100\,ms time step is chosen to allow for a further time span while also keeping the training time of the system relatively short. Also, We averaged the results of the Cross-CCC function instead of computing its maximum, as it can be more easily backpropagated during the training of the system. 

The second model is composed of a simple linear layer whose objective is to predict the GS obtained by calculating the weighted average of the corrected annotations, using the acoustic features as input. To ensure that the generated GS is well synchronised with the corresponding acoustic features, we used as loss function the CCC between the output of this model and the generated GS. 
We needed to use a linear layer here, i.e. without any usage of contextual information, in order not to compromise correction of the annotations in time by the first model. 

Because reliability of the annotation can vary by annotator, we performed a weighted average of the corrected annotations to produce the GS. This step is realised by a trainable weight vector that goes through a Softmax layer to return a probability vector, which is finally multiplied with the corrected annotations to obtain the GS. 

Since the two loss functions are jointly learnt, the B-GRU model ideally produces corrections to the annotations that match what can be linearly inferred from the features, and hence provides dynamic time-alignment of the annotations with the corresponding acoustic events. The loss function that is used for jointly training our system is defined as follows:

\begin{equation}
  \mathcal{L} = [1- \frac{1}{N}\sum_{n=1}^{N}{CrossCCC(.,.)_n}] \\+ [1-CCC(.,.)]
  \label{equ:Loss}
\end{equation}
where $N$ is the number of annotations, $CrossCCC(.,.)_n$ the averaged of the CCCs computed between the original $n$th annotations and the corrected annotations for different delays, and $CCC(.,.)$ is computed between the output of the linear model and the targeted GS with the following equation~\cite{li89}:
 \begin{equation}
   CCC(x,y) = \frac{2 \rho \sigma_x \sigma_y}{\sigma_x^2 + \sigma_y^2 + (\mu_x - \mu_y)^2}
   \label{CCC}
 \end{equation}
 where $\rho$ is the (Pearson's) correlation coefficient, $\mu_x$ and $\mu_y$ are the means for the vectors $x$ and $y$ respectively and their variances are defined as $\sigma_x$ and $\sigma_y$. And cross-CCC is computed through the following equation:
  \begin{equation}
   Cross-CCC(x(t),y(t)) = \frac{1}{N}\sum_{n=1}^{N}{CCC(x(t),y(t-T(n)))}
   \label{CrossCCC}
 \end{equation}
 where $T$ is a temporal series in seconds (here it is defined as $T=\{0, 0.1, 0.2, 0.3, \ldots, 9.9, 10\}$), $N$ is the total number of elements in $T$, $x$ is the original signal and $y(t-T(n))$ is the shifted version of $y(t)$ in time.

\begin{table}[t]
  \caption{Details of RECOLA and SEWA data sets used for experimental evaluations. }
  \label{tab:corpusSum}
  \centering
  \begin{tabular}{ l|c|c }
    \hline
    Characteristics & RECOLA & SEWA \\
     \hline
     \hline
    \# Female / Male & 27 / 19 & 103 / 101 \\
    \hline
    \# Annotators & 6 & 5 \\
    \hline
    Duration & 230 minutes & 510 minutes \\
    \hline
    Culture(s) & French & \begin{tabular}[x]{@{}c@{}}Chinese, \\German, \\Hungarian\end{tabular}  \\
    \hline
    \begin{tabular}[x]{@{}c@{}}\begin{tabular}[x]{@{}c@{}}Partitioning\\ Training-Dev-Test\end{tabular}\end{tabular} & 16-15-15 & \begin{tabular}[x]{@{}c@{}}\begin{tabular}[x]{@{}c@{}}GE: 34-14-16\\ HU: 34-14-18\\CN: 30-20-20\end{tabular}\end{tabular} \\
    \hline
  \end{tabular}
\end{table}
 
\section{Experiments}
\label{sec:expResults}

In this section, we describe the acoustic features that were extracted from the speech signals, the data sets we exploited, and the emotion prediction systems along the Sinc layer we used for experimental evaluations. Training and testing conditions for the models are also given, as well as the hyperparameter that was optimized.


\subsection{Features}
We extracted as acoustic descriptors the first 40 log Mel-filterbank (MFB) coefficients from the audio signal using the \textsc{python\_speech\_features}\footnote{https://github.com/jameslyons/python\_speech\_features} library, as they have proven to perform well with different models in emotion prediction tasks~\cite{le2017discretized, khorram2017capturing, albadawy2018joint, khorram2019jointly}. The feature extraction is realised on a 25\,ms window that is shifted forward in time each 10\,ms. All features are standardized using mean and variance obtained on the training partition of each data set, which makes them to have a zero mean and unit variance on this partition. 

\subsection{Data sets}
RECOLA~\cite{ringeval2013introducing} and SEWA~\cite{kossaifi2019sewa} data sets were used for our experiments. A summary of these corpora is given in Table \ref{tab:corpusSum}. RECOLA is a well-known corpus for benchmarking emotion recognition systems~\cite{Valstar16-A2D, Ringeval18-A2W}. It contains 46 audiovisual recordings of spontaneous interactions between French-speaking subjects who solved a collaborative task under remote condition. The recordings were captured in noise-free conditions with the same recording equipment. They were cut at five minutes, and further annotated in dimensions for both arousal and valence by six annotators with a sampling rate of 25\,Hz. The data set provides a GS that is computed as a consensus between the annotators~\cite{Valstar16-A2D}, which we will refer to as the baseline for our experiments. As a smaller version of this data set containing 27 subjects was used for a GS generation challenge~\cite{Ringeval18-A2W}, we also performed experiments with those 27 subjects in addition to the version with 46 subjects, to make fair comparisons of our results with the state-of-the-art. 

SEWA is a more recent corpus that consists of audiovisual recordings of spontaneous interactions between subjects for several different cultures. The data were recorded in-the-wild, i.e. via a video chat platform the webcams and microphones of each participant at various locations. The conversations were about an advertisement and are from 47 seconds to three minutes. Expressions of apparent emotion were annotated in the dimensions of arousal, valence and (dis)liking intensity (not used in this paper as the associated cues are mostly conveyed by linguistic information) with a sampling rate of 10\,Hz. We used the German (GE), Hungarian (HU) and Chinese (CN) cultures with the exact same partitioning as defined in the AVEC 2019 Challenge~\cite{ringeval2019avec}, with the exception of the Chinese culture that was solely used for testing, and which is partitioned here with the same rules as those used for the other cultures. The data set also provides a GS generated as a consensus between the annotators. 

To have the same length for both the features and annotations, which make the learning of the emotion prediction models easier, we linearly re-sampled all the annotations for all data sets to 100\,Hz, which is the sampling frequency of the audio features.

\subsection{Emotion prediction systems}
\begin{figure}[t]
  \centering
  \includegraphics[width=.75\linewidth]{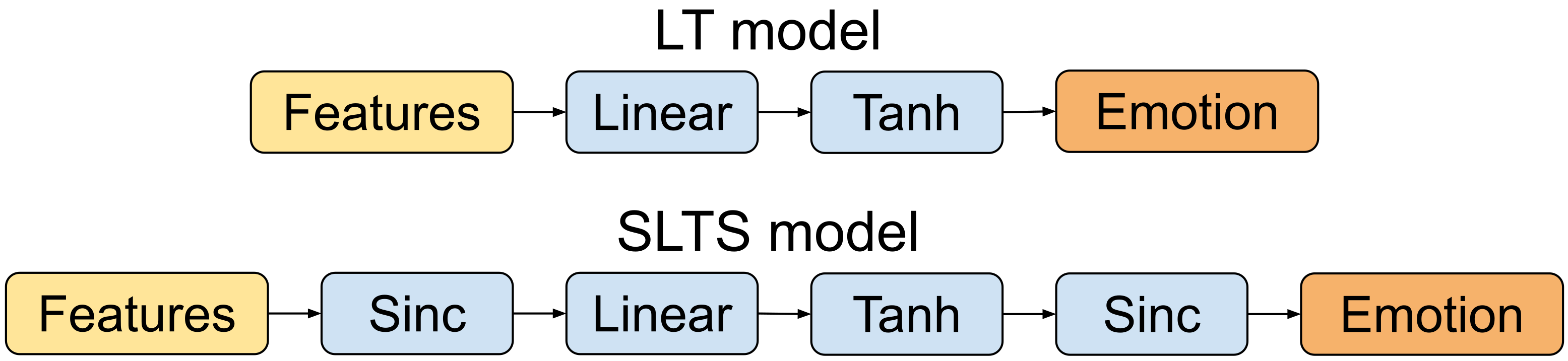}
  \caption{Flowcharts of the emotion prediction systems.}
  \label{fig:emoPredModels}
\end{figure}

In order to evaluate the ability of our system to synchronise annotations with features, we based our emotion prediction system on a simple linear layer that only maps a given input to one output, followed by a tangent hyperbolic function to map the output to the annotations' range, which is within $[-1,+1]$. We used a simple and light-weight model that does not exploit -- potentially long-range -- contextual dependencies between features and annotations, as recurrent architectures are not necessary to achieve competitive results for time-continuous emotion recognition with NNs~\cite{Schmitt19-CER}.
Benefits of the Sinc layer in emotion recognition from speech are further evaluated by incorporating it in the emotion prediction system. Flowcharts of the two developed emotion prediction systems, i.e. Linear-Tanh (LT), and Sinc-Linear-Tanh-Sinc (SLTS), can be found in Figure \ref{fig:emoPredModels}.

\begin{figure}[t]
  \centering
  \includegraphics[width=\linewidth]{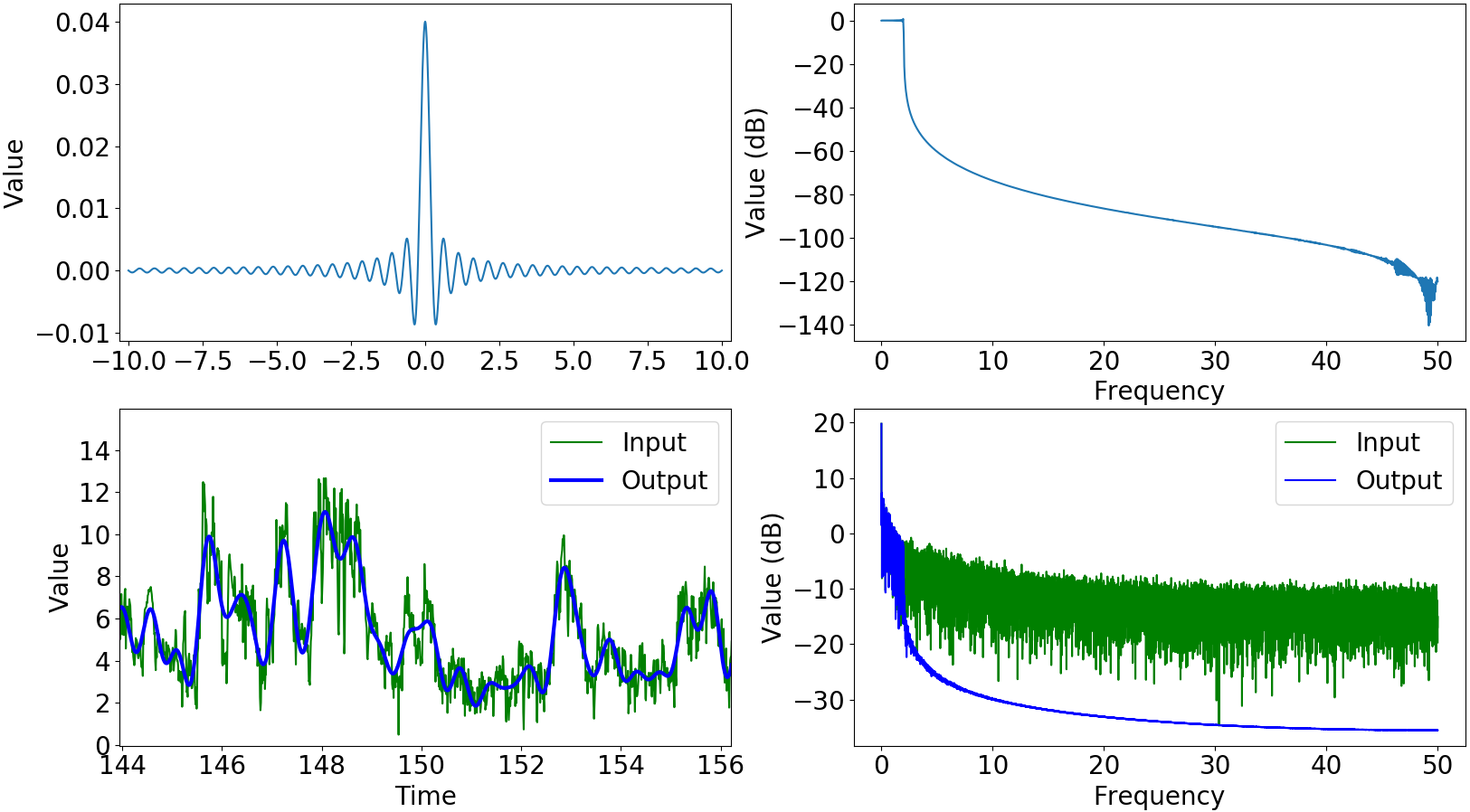}
  \caption{Effect of the Sinc layer with $f_c$=2\,Hz on the first MFB feature for subject "train\_01" of RECOLA data set. Top-left: Sinc function in time domain. 
  Top-right: frequency response of the Sinc function with a clear cut at $f_c$. Bottom-left: smoothing effect of the Sinc layer on the first MFB feature. Bottom-right: spectrum of the input and output of the Sinc layer.}
  
  \label{fig:sincLayer}
\end{figure}

\subsection{Sinc Layer}
The Sinc layer is composed of a convolutional layer with the kernel defined as the Sinc function:

\begin{equation}
  K_{sinc}(n) = \frac{Sin(2 \pi n f_c)}{\pi n f_s}
  \label{equ:Sinc}
\end{equation}
where $f_s$ is the sampling frequency (here $f_s$=100\,Hz), and $f_c$ the cutoff frequency, which is a trainable parameter in our case. 
The kernel size here is set to 20 seconds ($20*f_s$) in order to cover relatively low cutoff frequencies.
Using the Sinc layer on a signal acts as an ideal low-pass filter, which has a smoothing effect depending on the $f_c$. We initialised this cutoff frequency to $f_s/1000$ with the idea to let the system start the training with a low resolution signal, and go towards a more detailed representation if needed, i.e. for reaching a lower loss. We show the results of this Sinc layer when applied on the input data in Figure \ref{fig:sincLayer}.

\subsection{Training and Testing}
Training of the emotion prediction models was achieved with Adam optimizer with the initial learning rate set to .001. The maximum number of epochs was set to 250, and training was stopped if no improvement on the development set was observed after ten epochs. Learning was performed with a loss function based on CCC~\cite{weninger2016discriminatively}, which was computed locally, i.e. for each sequence. Reported CCCs for all the experiments are calculated by averaging the values obtained over subjects on the test partition. In order to ensure the results are both reliable and reproducible, we ran each experiment five times and report mean and variance over the runs. All the experiments were done with Pytorch~\cite{paszke2019pytorch} version 1.3.1 with seeds set to zero manually. The OS was Ububtu 18.04.3 LTS and the computer was equipped with an Nvidia Quadro RTX 4000 with 8 Giga-bytes of memory, CUDA version 10.0. 

\subsection{Hyper-parameter}
We optimised the hidden size of the B-GRU model used for annotation correction in the GS generator model. We experimented with hidden size $\in \{8, 16, 32, 64, 128, 256, 512\}$ on RECOLA data set and the lowest loss was achieved by hidden size of 128 for arousal and 256 for valence, which were used to generate the GS for all data sets.

\begin{figure*}[t]
  \centering
  \includegraphics[width=\linewidth]{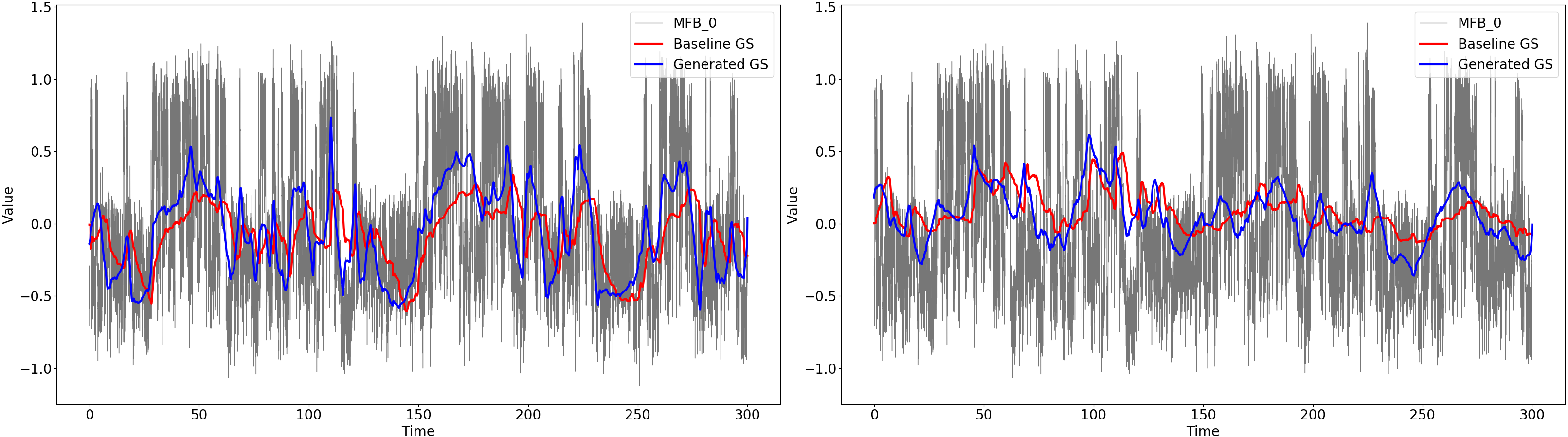}
  \caption{Comparisons of the alignment of the original and generated GS with the first MFB feature (divided by 2 for better visualisation) for arousal (left) and valence (right) on RECOLA data set for the subject "dev\_01".}
  
  \label{fig:arousalExample}
\end{figure*}

\section{Results}
\label{sec:Res}
In this section, we detail and discuss the qualitative and quantitative results obtained from the analysis of the generated GS.

\begin{table}[t]
  \caption{Statistics of the inter-annotator agreement (pair-wise Cronbach's alpha) obtained on each recording of RECOLA and SEWA data sets; [mean (variance)]; 'Baseline' refers to the original annotations; 'Generated' refers to the annotations corrected with our system.}
  \label{tab:cronbachAlphas}
  \centering
  \begin{tabular}{ l|c|c }
    \multicolumn{3}{c}{\textbf{Arousal}}\\
    \hline
    Data set & Baseline & Generated \\
     \hline
    RECOLA & .48 (.05) & \textbf{.53 (.05)} \\
    SEWA CN & .35 (.09) & \textbf{.43 (.16)} \\
    SEWA GE & \textbf{.25 (.07)} & .24 (.17) \\
    SEWA HU & \textbf{.11 (.15)} & .05 (.42) \\

     \hline
    \noalign{\vskip 2mm}
     
    \multicolumn{3}{c}{\textbf{Valence}}\\
    \hline
    Data set & Baseline & Generated \\
     \hline
    RECOLA & \textbf{.51 (.06)} & .51 (.09) \\
    SEWA CN & .35 (.09) & \textbf{.40 (.17)} \\
    SEWA GE & .38 (.09) & \textbf{.47 (.12)} \\
    SEWA HU & \textbf{.21 (.22)} & .14 (.77) \\

     \hline
  \end{tabular}
\end{table}
\subsection{Time-alignment with acoustic features}
Visual comparisons of the alignment of the baseline GS and that obtained with our system with the acoustic features are given in Figure \ref{fig:arousalExample}. The plot shows that the generated GS, compared to the baseline GS, is corrected both in time and value with respect to the corresponding acoustic events present in the speech. We further computed the Pearson's correlation coefficient between the first MFB feature (MFB$_0$) and the GS and averaged the results over all subjects of the RECOLA data set. Results show that the correlation increases from .221 to .484 on arousal when corrections are applied to the GS, and from .069 to .283 on valence.

\begin{figure}[t]
  \centering
  \includegraphics[width=.75\linewidth]{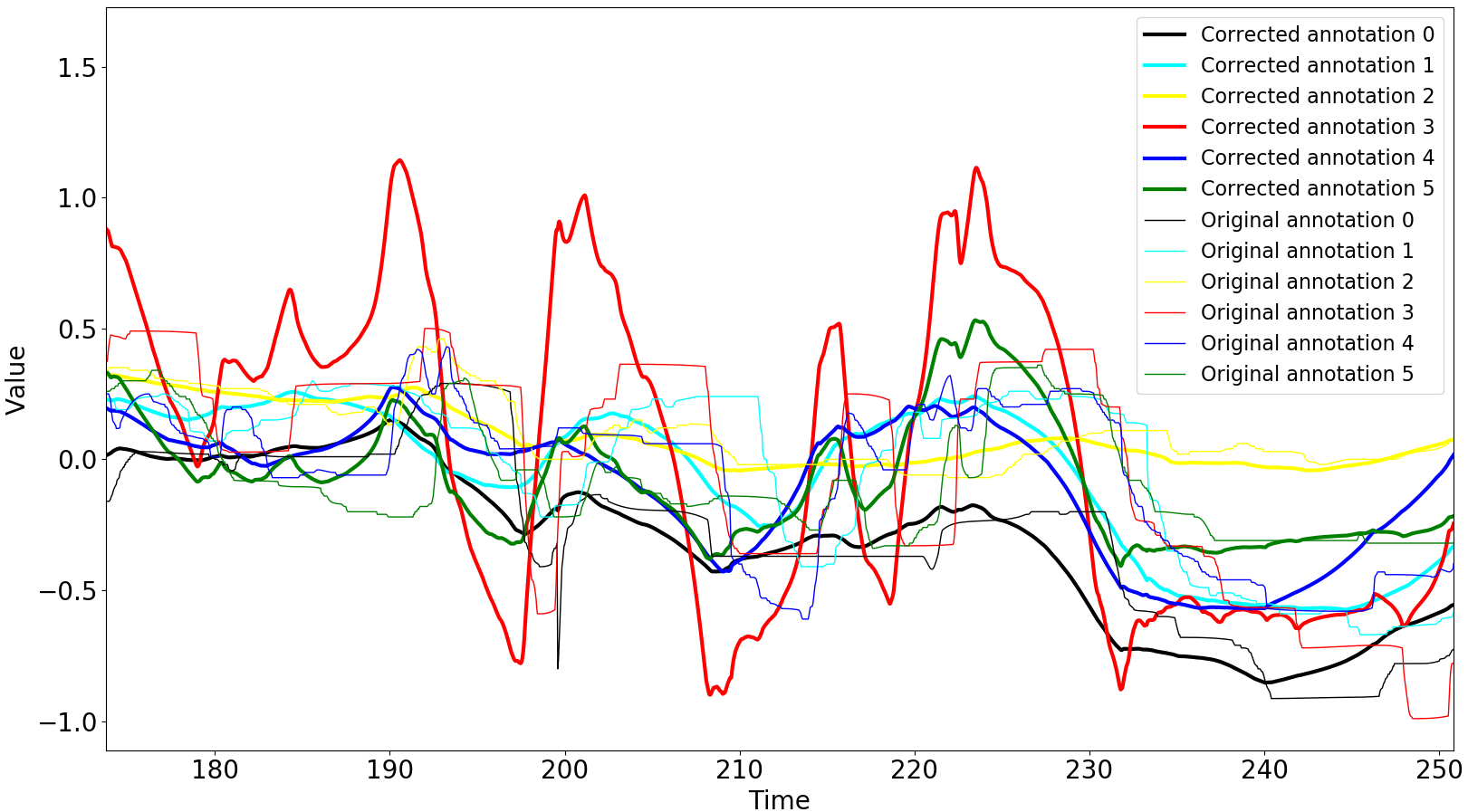}
  \caption{Comparison of the individual annotations before and after the B-GRU model, which is designed for preserving the original shape of the annotations when applying corrections to them. Thick lines correspond to the corrected annotation of the thin lines with the same color. The plot is for arousal and represents subject "dev\_01" of RECOLA data set.}
  \label{fig:CompAnnotsArousal}
\end{figure}

\subsection{Preservation of the original shape}
To ensure that the B-GRU model correcting the annotations does not distort them, we used a specific loss function (c.f. equation \ref{equ:Loss}) that has the objective of preserving the overall shape of the original annotations. We show in Figure \ref{fig:CompAnnotsArousal} that such information is indeed well preserved when our system corrects the annotations. 

\subsection{Inter-annotator agreement}
We report comparisons of the inter-annotator agreement obtained on the original and corrected annotations for RECOLA and SEWA data sets in Table \ref{tab:cronbachAlphas}.
The annotations of emotion being continuous in value, we computed the pair-wise Cronbach's alpha~\cite{cronbach1951coefficient} as measure of inter-annotator agreement on each recording. We used the Fisher r-to-z transform to perform statistical comparisons of the inter-annotator agreement with a one-tailed test. Results show that the inter-annotator agreement is either preserved or significantly increased ($p<.005$) when using our system for all cases except for the Hungarian culture, where the reported inter-annotator agreement is the lowest. This result suggests that a minimum of agreement between the annotators is required to perform time-alignment of the annotations with the corresponding acoustic features.

\subsection{Impact of the Sinc layer}

\begin{figure}[t]
  \centering
  \includegraphics[width=.75\linewidth]{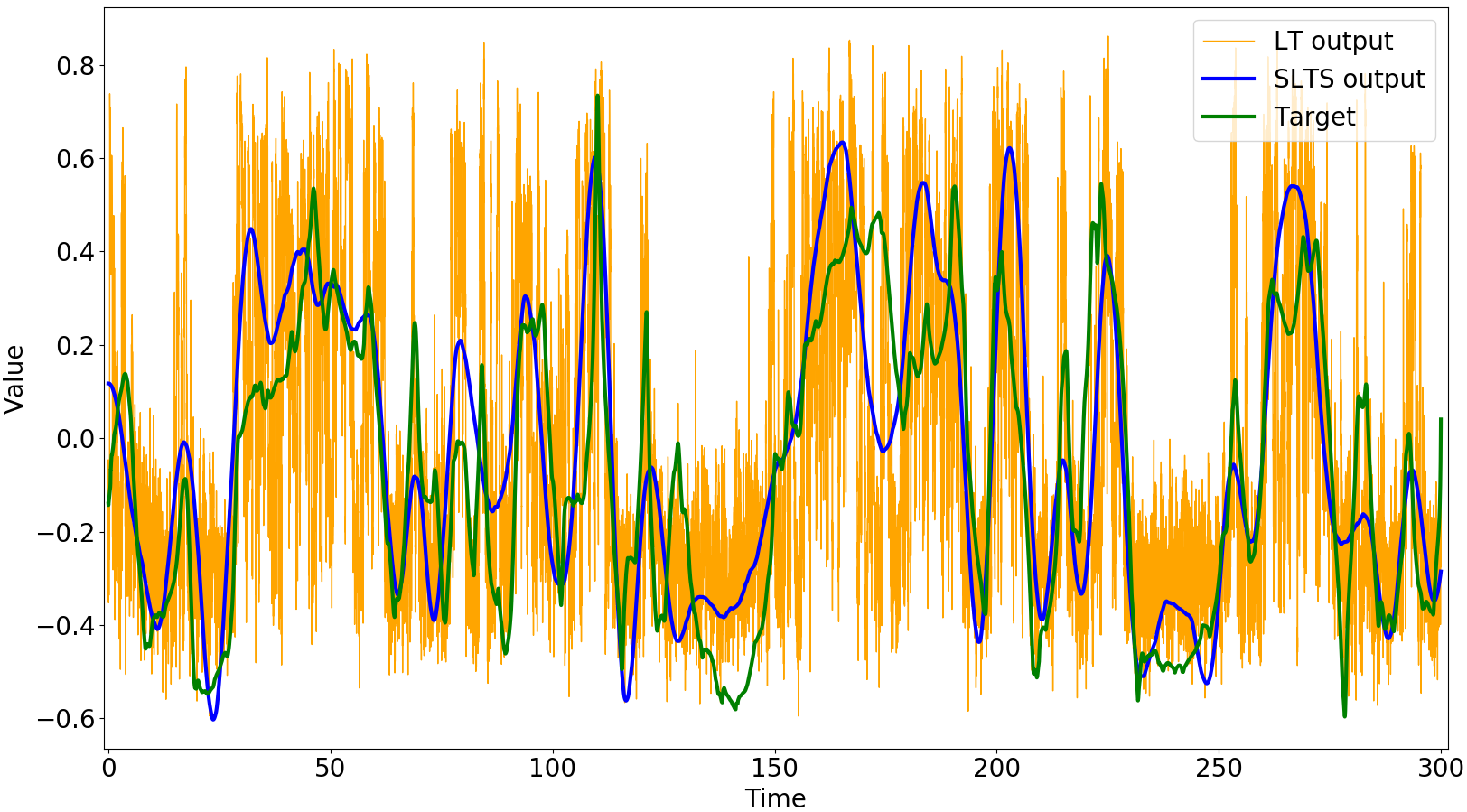}
  \caption{Predictions of arousal (target) obtained with the LT model, which includes a linear layer followed by hyperbolic tangent function, and the SLTS model, which additionally make use of the Sinc layer. The plot is for subject "dev\_01" on RECOLA data set.}
  \label{fig:compModelsOutputs}
\end{figure}

In Figure \ref{fig:compModelsOutputs}, we show the impact of the Sinc layer by comparing predictions obtained with the two emotion prediction systems along with the GS. Results show that the LT model returns predictions that contain very high frequencies, with strong variations over time, whereas the SLTS model provides predictions that have smooth trajectories.

We report the cutoff frequencies $f_c$ that were learned by the Sinc layers for the features and the predictions in Table \ref{tab:sincFcsRecola}. Results show that the details of the acoustic features seem to be useful only for the recognition of arousal in noise-free conditions, as the obtained cutoff frequencies are very low for all other cases.
Regarding the learnt cutoff frequencies for predictions, the results vary greatly depending on the dimension processed: whereas arousal does not need to be detailed precisely to reach the best performance, a common cutoff frequency of $f_s/2$ is obtained for valence, meaning that all details present in the GS were found useful for its automatic recognition.
One may further note that the Hungarian culture showed the lowest $f_c$ value in most of cases, which is in agreement with the results reported in Table \ref{tab:cronbachAlphas}.

\subsection{Performance in affect sensing}
We present the performance obtained by the two emotion prediction systems on the baseline and generated GS, for RECOLA and SEWA data sets, in Table \ref{tab:expComp}. 
Results show that the generated GS is more reliable to predict arousal or valence from speech, either in noise-free conditions or for data captured in-the-wild, and that the use of the Sinc layer always resulted in higher performance, except for the valence on the Hungarian culture. 

\begin{table}[t]
   \caption{Cutoff frequencies ($f_c$) learnt by the Sinc layer for the features and the predictions on RECOLA and SEWA data sets. We report only the mean over five different runs as the variance was zero for most of the cases.}
   \label{tab:sincFcsRecola}
   \centering
   \begin{tabular}{ l|c|c|c }
     \multicolumn{4}{c}{\textbf{Arousal}}\\
     \hline
     GS & Data set & $f_c$ Feature & $f_c$ Prediction \\
      \hline
     Baseline & RECOLA & 30.7 & 0.05 \\
     Baseline & SEWA CN & 0.38 & 0.04 \\
     Baseline & SEWA GE & 12.1 & 0.05 \\
     Baseline & SEWA HU & 0.02 & 0.02 \\
     
     \hline
     Generated & RECOLA & 1.45 & 0.10 \\
     Generated & SEWA CN & 0.02 & 0.60 \\
     Generated & SEWA GE & 0.16 & 0.15 \\
     Generated & SEWA HU & 0.01 & 0.60 \\
     
      \hline
     \noalign{\vskip 2mm}
     
     \multicolumn{4}{c}{\textbf{Valence}}\\
     \hline
     target GS & Data set & $f_c$ Feature & $f_c$ Prediction \\
      \hline
     Baseline & RECOLA & 0.03 & 23.26 \\
     Baseline & SEWA GE & 0.07 & 49.62 \\
     Baseline & SEWA HU & 0.03 & 49.82 \\
     Baseline & SEWA CN & 0.09 & 49.96 \\
     \hline
     Generated & RECOLA & 0.07 & 49.98 \\
     Generated & SEWA GE & 0.05 & 49.98 \\
     Generated & SEWA HU & 0.08 & 49.97 \\
     Generated & SEWA CN & 0.02 & 48.96 \\

      \hline
   \end{tabular}
 \end{table}

\begin{table*}[t]
  \caption{Performance as measured by CCC obtained by the two emotion prediction systems (LT, and SLTS) on the baseline and the generated GS of RECOLA and SEWA data sets. The reported values are only the mean of CCCs over five runs, as the variance was zero when rounded to three floating point. To ease overall comparison, we provide an average over the different data sets for each model and target, which is presented in the last column.}
  \label{tab:expComp}
  \centering
  \begin{tabular}{ l|c|c|c|c|c|c }
    \multicolumn{7}{c}{\textbf{Arousal}}\\
    \hline
     
     Target GS & Model & RECOLA--French & SEWA--Chinese & SEWA--German & SEWA--Hungarian & All \\
     \hline
    Baseline & LT & .190 & .000 & .102 & .000 & .073 \\
    Baseline & SLTS & .528 & .082 & .249 & \textbf{.130} & .247 \\
    \hline
    Generated & LT & .482 & .054 & .367 & .071 & .243 \\
    Generated & SLTS & \textbf{.720}  & \textbf{.120} & \textbf{.450} & .123 & \textbf{.353} \\
     \hline
     
    \multicolumn{7}{c}{\textbf{Valence}}\\
    \hline
    Target GS & Model & RECOLA--French & SEWA--Chinese & SEWA--German & SEWA--Hungarian & All \\
    \hline
    Baseline & LT & .074 & .051 & .037 & .005 & .042 \\
    Baseline & SLTS & .289 & \textbf{.137} & \textbf{.134} & .000 & .140 \\
    \hline
    Generated & LT & .248 & .049 & .001 & .000 & .074 \\
    Generated & SLTS & \textbf{.538} & .117 & .119 & \textbf{.066} & \textbf{.210} \\
    \hline
  \end{tabular}
\end{table*}

\begin{table*}[t]
  \caption{Comparison of the results obtained in this study with those of the state-of-the-art on RECOLA and SEWA data sets. Reported results are on the test partition and for audio only features.}
  \label{tab:compStateOfTheArt}
  \centering
  
  \begin{tabular}{ l|c|c|c }
    \multicolumn{4}{c}{\textbf{RECOLA}}\\
        \hline
        Reference & \# Subjects & Arousal & Valence \\
        \hline
        \hline
        Original GS + LSTM~\cite{ringeval2015prediction} & 46 & .788 & .412 \\
        Original GS + Support Vector Regression~\cite{schmitt2016border} & 46 & .738 & .430 \\
        Original GS + Support Vector Regression~\cite{han2018bags} & 46 & \textbf{.750} & .465 \\
        \hline
        Generated GS + SLTS (this paper) & 46 & .720 & \textbf{.538} \\
        \hline
        \hline
        Original GS + Support Vector Regression~\cite{Ringeval18-A2W} & 27 & .651 & .346 \\
        Original GS + GRUs~\cite{zhang2018dynamic} & 27 & \textbf{.699} & .449 \\
        Original GS + CNNs~\cite{khorram2019jointly} & 27 & .688 & \textbf{.492} \\
        \hline
        Generated GS + SLTS (this paper) & 27 & .513 & .427 \\
    \hline
  \end{tabular}
  
  \vskip 2mm
  
  \begin{tabular}{ l|c|c|c|c }
    \multicolumn{5}{c}{\textbf{SEWA}}\\
        \hline
        \multicolumn{1}{c}{\textbf{Target emotion}} & \multicolumn{2}{c}{
        \textbf{\rule[2 pt]{30 pt}{0.5 pt}} \textbf{Arousal} \textbf{\rule[2 pt]{30 pt}{0.5 pt}}
        } & \multicolumn{2}{c}{\textbf{\rule[2 pt]{30 pt}{0.5 pt}} \textbf{Valence} \textbf{\rule[2 pt]{30 pt}{0.5 pt}} }\\
        \hline
        Reference & German & Hungarian & German & Hungarian \\
        \hline
        \hline
        Original GS + LSTM~\cite{ringeval2019avec} & .296 & \textbf{.160} & \textbf{.288} & -0.019 \\
        \hline
        Generated GS + SLTS (this paper) & \textbf{.450} & .123 & .119 & \textbf{.066} \\
    \hline
  \end{tabular}
  
\end{table*}

\subsection{Comparison with state-of-the-art}
We compare the results obtained in this study with those of the state-of-the-art in Table \ref{tab:compStateOfTheArt}. In order to make the comparisons fair on RECOLA data set, we distinguish its two different versions and present our results accordingly: one containing 27 subjects, and another with all 46 subjects. Regarding the results reported on SEWA data set, we used as a reference the performance obtained with the MFCC features in the baseline system developed for AVEC 2019 with intra-cultural models~\cite{ringeval2019avec}, for which performance is not reported on the Chinese culture. 
Results show that the generated GS can perform similar to the state-of-the-art, while using a relatively simpler network, and that improvement can be further reported, especially on arousal for recordings captured in-the-wild, and valence for noise-free conditions.

\subsection{Discussion}
\label{sec:discussion}

As there is no ground-truth in the description of human apparent emotion, defining numerical attributes that describe expressions of affect in a way that they can be easily recognised from speech with machine learning methods is a challenging task. Indeed, when humans are confronted with fine-grained annotation of emotion attributes with time-continuous scales, such as arousal and valence, many discrepancies can be found between annotators, either in time or value, adding noise to the definition of targets used for emotion recognition. In this work, we show that audio signals can be used to correct such inconsistencies, and that the resulting GS can provide a higher inter-annotator agreement, and better performance in its automatic recognition from speech. However, it should be noted that because the annotations were performed on the audiovisual recordings, visual information was also available to the annotator to report on the apparent expressions of emotion. As emotion is inherently multimodal, our method needs to be further investigated using facial information in addition to acoustic descriptors. The use of peripheral physiological signals could also be investigated, as well as recordings of the annotators when performing the annotation.

\section{Conclusions}
\label{sec:conclusion}

In this paper, we proposed a system that can compensate for the inconsistencies found in continuous emotion annotations with constraints driven from audio features. We showed that the generated GS, which preserves the details of the original annotations, can provide a significant increase in both inter-annotator agreement and performance obtained in its automatic recognition for different cultures and contexts. 

We also investigated the interest of a Sinc layer acting as a low-pass filter with a cutoff frequency that can be learnt for features and labels. The Sinc layer initialised from a low value can help the model follow curriculum learning. The results shows that the method always improved performance and furthermore provided interpretable results.

Future work will involve exploiting more sophisticated descriptors of speech, such as self-supervised representations, as well as facial descriptors, to perform annotation corrections. We are also interested in modeling each individual annotators' reaction during the annotation of time-continuous attributes of emotion using their own audiovisual recordings.

\section*{Acknowledgment}
The research leading to these results has received funding from the Association Nationale de la Recherche et de la Technologie (ANRT), under grant agreements No. 2019/0729 (Wellbot project).

\bibliographystyle{unsrtnat}
\bibliography{references}  

\begin{thebibliography}{34}
\providecommand{\natexlab}[1]{#1}
\providecommand{\url}[1]{\texttt{#1}}
\expandafter\ifx\csname urlstyle\endcsname\relax
  \providecommand{\doi}[1]{doi: #1}\else
  \providecommand{\doi}{doi: \begingroup \urlstyle{rm}\Url}\fi

\bibitem[Tsatsou et~al.(2018)Tsatsou, Pomazanskyi, Hortal, Spyrou, Leligou, and
  Asteriadis]{Tsatsou18-ALB}
Dorothea Tsatsou, Andrew Pomazanskyi, Enrique Hortal, Evaggelos Spyrou,
  Helen~C. Leligou, and Stylianos Asteriadis.
\newblock Adaptive learning based on affect sensing.
\newblock In \emph{Proceedings of the International Conference on Artificial
  Intelligence in Education (AIED)}, volume~6, pages 475--479, London, UK,
  2018. Springer.
\newblock LNCS, volume 10948.

\bibitem[Cummins et~al.(2015)Cummins, Scherer, Krajewski, Schnieder, Epps, and
  Quatieri]{cummins2015review}
Nicholas Cummins, Stefan Scherer, Jarek Krajewski, Sebastian Schnieder, Julien
  Epps, and Thomas~F Quatieri.
\newblock A review of depression and suicide risk assessment using speech
  analysis.
\newblock \emph{Speech Communication}, 71:\penalty0 10--49, 2015.

\bibitem[Cosentino et~al.(2018)Cosentino, Randria, Lin, Pellegrini, Sessa, and
  Takanishi]{cosentino2018group}
Sarah Cosentino, Estelle~IS Randria, Jia-Yeu Lin, Thomas Pellegrini, Salvatore
  Sessa, and Atsuo Takanishi.
\newblock Group emotion recognition strategies for entertainment robots.
\newblock In \emph{International Conference on Intelligent Robots and Systems
  (IROS)}, pages 813--818, Madrid, Spain, 2018. IEEE, RSJ.

\bibitem[Ekman(1993)]{Ekman93-FEA}
Paul Ekman.
\newblock Facial expression and emotion.
\newblock \emph{American Psychologist}, 28\penalty0 (4):\penalty0 384--392,
  1993.

\bibitem[Ekman and Cordaro(2011)]{Ekman11-WIM}
Paul Ekman and Daniel Cordaro.
\newblock What is meant by calling emotions basic.
\newblock \emph{Emotion Review}, 3\penalty0 (4):\penalty0 364--370, 2011.

\bibitem[Russel(1980)]{Russell80-ACM}
James~A. Russel.
\newblock A circumplex model of affect.
\newblock \emph{Journal of Personality and Social Psychology}, 39\penalty0
  (6):\penalty0 1161--1178, 1980.

\bibitem[Wehrle and Scherer(2001)]{Wehrle01-TCM}
Thomas Wehrle and Klaus Scherer.
\newblock Towards computational modeling of appraisal theories.
\newblock In \emph{Appraisal Processes in Emotion: Theory, Methods, Research},
  pages 350--365. New York: Oxford University Press, 2001.

\bibitem[Schuller et~al.(2012)Schuller, Valstar, Eyben, Cowie, and
  Pantic]{schuller2012avec}
Bj\"orn Schuller, Michel Valstar, Florian Eyben, Roddy Cowie, and Maja Pantic.
\newblock {AVEC 2012 -- The continuous Audio/Visual Emotion Challenge}.
\newblock In \emph{{Proceedings of the 14th International Conference on
  Multimodal Interaction (ICMI)}}, pages 449--456, Santa Monica (CA), USA,
  2012. ACM.

\bibitem[Soroosh and Busso(2013)]{Soroosh13-AAC}
Mariooryad Soroosh and Carlos Busso.
\newblock {Analysis and Compensation of the Reaction Lag of Evaluators in
  Continuous Emotional Annotations}.
\newblock In \emph{Proceedings of the 5th biannual International Conference on
  Affective Computing and Intelligent Interaction (ACII)}, pages 85--90,
  Geneva, Switzerland, 2013. IEEE.

\bibitem[Ringeval et~al.(2018)Ringeval, Schuller, Valstar, Cowie, Kaya,
  Schmitt, Amiriparian, Cummins, Lalanne, Michaud, Ciftci, G\"ulec, Salah, and
  Pantic]{Ringeval18-A2W}
Fabien Ringeval, Bj\"orn Schuller, Michel Valstar, Roddy Cowie, Heysem Kaya,
  Maximilian Schmitt, Shahin Amiriparian, Nicholas Cummins, Dennis Lalanne,
  Adrien Michaud, Elvan Ciftci, H\"useyin G\"ulec, Albert~Ali Salah, and Maja
  Pantic.
\newblock {AVEC 2018 Workshop and Challenge: Bipolar Disorder and
  Cross-Cultural Affect Recognition}.
\newblock In \emph{{Proceedings of the 8th International Workshop on
  Audio/Visual Emotion Challenge (AVEC'18)}}, pages 3--13, Seoul, South Korea,
  2018. ACM.

\bibitem[Khorram et~al.(2019)Khorram, McInnis, and Provost]{khorram2019jointly}
Soheil Khorram, Melvin McInnis, and Emily~Mower Provost.
\newblock Jointly aligning and predicting continuous emotion annotations.
\newblock \emph{IEEE Transactions on Affective Computing}, 2019.
\newblock (Early access).

\bibitem[Nicolle et~al.(2012)Nicolle, Rapp, Bailly, Prevost, and
  Chetouani]{nicolle2012robust}
J{\'e}r{\'e}mie Nicolle, Vincent Rapp, K{\'e}vin Bailly, Lionel Prevost, and
  Mohamed Chetouani.
\newblock Robust continuous prediction of human emotions using multiscale
  dynamic cues.
\newblock In \emph{Proceedings of the 14th ACM International Conference on
  Multimodal Interaction (ICMI'12)}, pages 501--508, Santa Monica (CA), USA,
  2012. ACM.

\bibitem[Mariooryad and Busso(2014)]{mariooryad2014correcting}
Soroosh Mariooryad and Carlos Busso.
\newblock Correcting time-continuous emotional labels by modeling the reaction
  lag of evaluators.
\newblock \emph{IEEE Transactions on Affective Computing}, 6\penalty0
  (2):\penalty0 97--108, 2014.

\bibitem[Khorram et~al.(2017)Khorram, Aldeneh, Dimitriadis, McInnis, and
  Provost]{khorram2017capturing}
Soheil Khorram, Zakaria Aldeneh, Dimitrios Dimitriadis, Melvin McInnis, and
  Emily~Mower Provost.
\newblock Capturing long-term temporal dependencies with convolutional networks
  for continuous emotion recognition.
\newblock \emph{arXiv preprint}, 2017.
\newblock ({arXiv}:1708.07050).

\bibitem[Bengio et~al.(2009)Bengio, Louradour, Collobert, and
  Weston]{bengio2009curriculum}
Yoshua Bengio, J{\'e}r{\^o}me Louradour, Ronan Collobert, and Jason Weston.
\newblock Curriculum learning.
\newblock In \emph{Proceedings of the 26th Annual International Conference on
  Machine Learning (ICML'09)}, pages 41--48, Montreal, Canada, 2009. ACM.

\bibitem[Li(1989)]{li89}
Lin Li.
\newblock A concordance correlation coefficient to evaluate reproducibility.
\newblock \emph{Biometrics}, 45\penalty0 (1):\penalty0 255--268, March 1989.

\bibitem[He et~al.(2015)He, Jiang, Yang, Pei, Wu, and Sahli]{he2015multimodal}
Lang He, Dongmei Jiang, Le~Yang, Ercheng Pei, Peng Wu, and Hichem Sahli.
\newblock Multimodal affective dimension prediction using deep bidirectional
  long short-term memory recurrent neural networks.
\newblock In \emph{Proceedings of the 5th International Workshop on
  Audio/Visual Emotion Challenge (AVEC'15)}, pages 73--80. ACM, 2015.

\bibitem[Booth et~al.(2018)Booth, Mundnich, and Narayanan]{booth2018fusing}
Brandon~M Booth, Karel Mundnich, and Shrikanth Narayanan.
\newblock Fusing annotations with majority vote triplet embeddings.
\newblock In \emph{Proceedings of the 2018 Audio/Visual Emotion Challenge and
  Workshop}, pages 83--89, Seoul, South Korea, 2018. ACM.

\bibitem[Ringeval et~al.(2015)Ringeval, Eyben, Kroupi, Yuce, Thiran, Ebrahimi,
  Lalanne, and Schuller]{ringeval2015prediction}
Fabien Ringeval, Florian Eyben, Eleni Kroupi, Anil Yuce, Jean-Philippe Thiran,
  Touradj Ebrahimi, Denis Lalanne, and Bj\"orn Schuller.
\newblock {Prediction of Asynchronous Dimensional Emotion Ratings from
  Audiovisual and Physiological Data}.
\newblock \emph{Pattern Recognition Letters}, 66:\penalty0 22--30, November
  2015.

\bibitem[Ravanelli and Bengio(2018{\natexlab{a}})]{ravanelli2018speaker}
Mirco Ravanelli and Yoshua Bengio.
\newblock Speaker recognition from raw waveform with sincnet.
\newblock In \emph{Proceedings of the IEEE Spoken Language Technology Workshop
  (SLT)}, pages 1021--1028, Athens, Greece, 2018{\natexlab{a}}. IEEE.

\bibitem[Ravanelli and Bengio(2018{\natexlab{b}})]{ravanelli2018interpretable}
Mirco Ravanelli and Yoshua Bengio.
\newblock Interpretable convolutional filters with sincnet.
\newblock \emph{arXiv preprint}, 2018{\natexlab{b}}.
\newblock ({arXiv}:1811.09725).

\bibitem[Le et~al.(2017)Le, Aldeneh, and Provost]{le2017discretized}
Duc Le, Zakaria Aldeneh, and Emily~Mower Provost.
\newblock Discretized continuous speech emotion recognition with multi-task
  deep recurrent neural network.
\newblock In \emph{Proceedings INTERSPEECH 2017, 18th Annual Conference of the
  International Speech Communication Association}, pages 1108--1112, Stockholm,
  Sweden, 2017.

\bibitem[AlBadawy and Kim(2018)]{albadawy2018joint}
Ehab~A AlBadawy and Yelin Kim.
\newblock Joint discrete and continuous emotion prediction using ensemble and
  end-to-end approaches.
\newblock In \emph{Proceedings of the 20th International Conference on
  Multimodal Interaction (ICMI'18)}, pages 366--375, Boulder (CO), USA, 2018.
  ACM.

\bibitem[Ringeval et~al.(2013)Ringeval, Sonderegger, Sauer, and
  Lalanne]{ringeval2013introducing}
Fabien Ringeval, Andreas Sonderegger, J\"urgen Sauer, and Denis Lalanne.
\newblock {Introducing the RECOLA Multimodal Corpus of Remote Collaborative and
  Affective Interactions}.
\newblock In \emph{Proceedings of the 2nd International Workshop on Emotion
  Representation, Analysis and Synthesis in Continuous Time and Space
  (EmoSPACE)}, Shanghai, China, 2013. IEEE.

\bibitem[Kossaifi et~al.(2019)Kossaifi, Walecki, Panagakis, Shen, Schmitt,
  Ringeval, Han, Pandit, Schuller, Star, Hajiyev, and Pantic]{kossaifi2019sewa}
Jean Kossaifi, Robert Walecki, Yannis Panagakis, Jie Shen, Maximilian Schmitt,
  Fabien Ringeval, Jing Han, Vedhas Pandit, Bj\"orn Schuller, Kam Star, Elnar
  Hajiyev, and Maja Pantic.
\newblock {SEWA DB: A Rich Database for Audio-Visual Emotion and Sentiment
  Research in the Wild}.
\newblock \emph{IEEE Transactions on Pattern Analysis and Machine
  Intelligence}, 41, 2019.
\newblock (Early access).

\bibitem[Valstar et~al.(2016)Valstar, Gratch, Schuller, Ringeval, Lalanne,
  {Torres Torres}, Scherer, Stratou, Cowie, and Pantic]{Valstar16-A2D}
Michel Valstar, Jonathan Gratch, Bj\"orn Schuller, Fabien Ringeval, Denis
  Lalanne, Mercedes {Torres Torres}, Stefan Scherer, Giota Stratou, Roddy
  Cowie, and Maja Pantic.
\newblock {AVEC 2016 -- Depression, Mood, and Emotion Recognition Workshop and
  Challenge}.
\newblock In \emph{{Proceedings of the 6th International Workshop on
  Audio/Visual Emotion Challenge (AVEC'16)}}, pages 3--10, Amsterdam, The
  Netherlands, 2016. ACM.

\bibitem[Ringeval et~al.(2019)Ringeval, Schuller, Valstar, Cummins, Cowie,
  Soleymani, Schmitt, Amiriparian, Messner, Tavabi, Song, Alisamir, Lui, Zhao,
  and Pantic]{ringeval2019avec}
Fabien Ringeval, Bj\"orn Schuller, Michel Valstar, Nicholas Cummins, Roddy
  Cowie, Mohammad Soleymani, Maximilian Schmitt, Shahin Amiriparian, Eva-Maria
  Messner, Leili Tavabi, Siyang Song, Sina Alisamir, Shuo Lui, Ziping Zhao, and
  Maja Pantic.
\newblock {AVEC 2019 Workshop and Challenge: State-of-Mind, Depression with AI,
  and Cross-Cultural Affect Recognition}.
\newblock In \emph{{Proceedings of the 9th International Workshop on
  Audio/Visual Emotion Challenge (AVEC'19)}}, pages 3--12, Nice, France, 2019.
  ACM.

\bibitem[Schmitt et~al.(2019)Schmitt, Cummins, and Schuller]{Schmitt19-CER}
Maximilian Schmitt, Nicholas Cummins, and Bj\"orn~W.\ Schuller.
\newblock {Continuous Emotion Recognition in Speech -- Do We Need Recurrence?}
\newblock In \emph{{Proceedings INTERSPEECH 2019, 20th Annual Conference of the
  International Speech Communication Association}}, pages 2808--2812, Graz,
  Austria, September 2019. ISCA, ISCA.

\bibitem[Weninger et~al.(2016)Weninger, Ringeval, Marchi, and
  Schuller]{weninger2016discriminatively}
Felix Weninger, Fabien Ringeval, Erik Marchi, and Bj\"orn Schuller.
\newblock {Discriminatively Trained Recurrent Neural Networks for Continuous
  Dimensional Emotion Recognition from Audio}.
\newblock In \emph{{Proceedings of the 25th International Joint Conference on
  Artificial Intelligence (IJCAI)}}, pages 2196--2202, New York City (NY), USA,
  2016. IJCAI/AAAI.

\bibitem[Paszke et~al.(2019)Paszke, Gross, Massa, Lerer, Bradbury, Chanan,
  Killeen, Lin, Gimelshein, Antiga, Lin, Gimelshein, and
  Antiga]{paszke2019pytorch}
Adam Paszke, Sam Gross, Francisco Massa, Adam Lerer, James Bradbury, Gregory
  Chanan, Trevor Killeen, Zeming Lin, Natalia Gimelshein, Luca Antiga, Zeming
  Lin, Natalia Gimelshein, and Luca Antiga.
\newblock Pytorch: An imperative style, high-performance deep learning library.
\newblock In \emph{Proceedings of the thirty-third Conference on Neural
  Information Processing Systems (NIPS)}, pages 8026--8037, Vancouver, Canada,
  2019. Neural Information Processing Systems Foundation.

\bibitem[Cronbach(1951)]{cronbach1951coefficient}
Lee~J Cronbach.
\newblock Coefficient alpha and the internal structure of tests.
\newblock \emph{Psychometrika}, 16\penalty0 (3):\penalty0 297--334, 1951.

\bibitem[Schmitt et~al.(2016)Schmitt, Ringeval, and
  Schuller]{schmitt2016border}
Maximilian Schmitt, Fabien Ringeval, and Bj\"orn Schuller.
\newblock {At the Border of Acoustics and Linguistics: Bag-of-Audio-Words for
  the Recognition of Emotions in Speech}.
\newblock In \emph{{Proceedings INTERSPEECH 2016, 17th Annual Conference of the
  International Speech Communication Association}}, pages 495--499, San
  Fransisco (CA), USA, 2016. ISCA.

\bibitem[Han et~al.(2018)Han, Zhang, Schmitt, Ren, Ringeval, and
  Schuller]{han2018bags}
Jing Han, Zixing Zhang, Maximilian Schmitt, Zhao Ren, Fabien Ringeval, and
  Bj\"orn Schuller.
\newblock {Bags in Bag: Generating Context-Aware Bags for Tracking Emotions
  from Speech}.
\newblock In \emph{{Proceedings INTERSPEECH 2018, 19th Annual Conference of the
  International Speech Communication Association}}, pages 3082--3086,
  Hyderabad, India, 2018. ISCA.

\bibitem[Zhang et~al.(2018)Zhang, Han, Coutinho, and
  Schuller]{zhang2018dynamic}
Zixing Zhang, Jing Han, Eduardo Coutinho, and Bj{\"o}rn Schuller.
\newblock Dynamic difficulty awareness training for continuous emotion
  prediction.
\newblock \emph{IEEE Transactions on Multimedia}, 21\penalty0 (5):\penalty0
  1289--1301, 2018.

\end{thebibliography}






\end{document}